\documentclass[3p,times,twocolumn]{elsarticle}

\usepackage{ecrc}

\usepackage{lineno}

\volume{00}

\firstpage{1}

\journalname{Nuclear Physics B Proceedings Supplement}

\runauth{Daniele Zanzi}

\jid{nppp}

\jnltitlelogo{Nuclear Physics B Proceedings Supplement}

\usepackage{amssymb}
\usepackage[figuresright]{rotating}

\usepackage{xspace}

\newcommand{\cp}{${\cal{CP}}$\xspace}

\begin{document}
%\linenumbers
\begin{frontmatter}

%% Title, authors and addresses

%% use the tnoteref command within \title for footnotes;
%% use the tnotetext command for the associated footnote;
%% use the fnref command within \author or \address for footnotes;
%% use the fntext command for the associated footnote;
%% use the corref command within \author for corresponding author footnotes;
%% use the cortext command for the associated footnote;
%% use the ead command for the email address,
%% and the form \ead[url] for the home page:
%%
%% \title{Title\tnoteref{label1}}
%% \tnotetext[label1]{}
%% \author{Name\corref{cor1}\fnref{label2}}
%% \ead{email address}
%% \ead[url]{home page}
%% \fntext[label2]{}
%% \cortext[cor1]{}
%% \address{Address\fnref{label3}}
%% \fntext[label3]{}

\dochead{}
%% Use \dochead if there is an article header, e.g. \dochead{Short communication}

\title{Measurement of the Higgs Boson Couplings and CP Structure Using Tau Leptons at the LHC}

%% use optional labels to link authors explicitly to addresses:
%% \author[label1,label2]{<author name>}
%% \address[label1]{<address>}
%% \address[label2]{<address>}

\author{Daniele Zanzi on behalf of the ATLAS and CMS Collaborations}

\address{ARC Centre of Excellence for Particle Physics at the Terascale, School of Physics, University of Melbourne, Victoria 3010, Australia}

\begin{abstract}
%% Text of abstract
Results on the $H\to\tau\tau$ measurements performed by the ATLAS and CMS collaborations with the $pp$ collision data collected at the LHC at $\sqrt{s}=7$ and 8 TeV are presented. These include a test of \cp invariance in the VBF Higgs boson production. Experimental challenges for the test of the \cp invariance in the $H\to\tau\tau$ decays are also reviewed.
\end{abstract}

\begin{keyword}
%% keywords here, in the form: keyword \sep keyword
LHC \sep ATLAS \sep CMS \sep Higgs \sep Tau \sep CP Violation
%% MSC codes here, in the form: \MSC code \sep code
%% or \MSC[2008] code \sep code (2000 is the default)

\end{keyword}

\end{frontmatter}

%%
%% Start line numbering here if you want
%%
% \linenumbers

%% main text
\section{Introduction}
\label{sec:intro}
%The discovery of a Higgs boson by the ATLAS and CMS experiments~\cite{Aad:2012tfa,Chatrchyan:2012xdj} at the LHC~\cite{Evans:2008zzb} offers a novel opportunity to search for new sources of \cp violation in the interactions of the Higgs boson with other Standard Model (SM) particles. These \cp-violating interactions could provide additional sources of \cp violation needed to explain the baryon asymmetry \cite{Bernreuther:2002uj}, the difference between the amounts of matter and antimatter observed in the Universe  \cite{Ade:2015xua}, that cannot be explained by the SM \cite{Huet:1994jb,Gavela:1993ts}. 
The search for \cp violation in the interactions of the newly discovered Higgs boson \cite{Aad:2012tfa,Chatrchyan:2012xdj} with the other Standard Model (SM) particles is motivated by the lack of sources of \cp violation to explain the baryon asymmetry observed in the Universe \cite{Bernreuther:2002uj,Ade:2015xua,Huet:1994jb,Gavela:1993ts}.
In the SM, no effect of \cp violation is expected at LO in the production or decay of the SM Higgs boson. Hence, an observation of \cp violation involving the observed Higgs boson  would be a strong sign of physics beyond the SM.

The $H\to\tau\tau$ final state is very powerful for studies of \cp invariance of the Higgs boson couplings. It is one of the most sensitive channels for the Vector Boson Fusion (VBF) Higgs boson production and it is the most sensitive for the Higgs boson decay into fermions. With $H\to\tau\tau$ events, it is possible to probe the \cp structure of both the $HVV$ couplings to gauge bosons in VBF events and also of the $Hff$ couplings to fermions. The $H\to\tau\tau$ channel is however challenging to analyse due to the presence of neutrinos, the high $Z\to\tau\tau$ production cross section, the difficult discrimination between hadronically decaying tau leptons ($\tau_{\rm had}$) from QCD jets, and the availability of only few and low-energy signatures to trigger on.

In this review, results on the $H\to\tau\tau$ measurements performed by the ATLAS \cite{Aad:2008zzm} and CMS \cite{Chatrchyan:2008aa} collaborations with the LHC Run-I data are presented. A first test of \cp invariance in the VBF production is also described, together with experimental challenges for the measurement of the \cp invariance in the $H\to\tau\tau$ decays.

\section{The $H\to\tau\tau$ Coupling Measurement}
\label{sec:Htt}
Evidence for the Higgs boson decays into tau leptons has been established by ATLAS \cite{Aad:2015vsa} and CMS \cite{Chatrchyan:2014nva} with the $pp$ collisions collected at the LHC at $\sqrt{s}=7$ and 8 TeV. The analyses performed by the two collaborations are similar in terms of event selections and background estimations. The main difference is the observable used for the signal extraction. The CMS analysis performs a combined shape fit of the reconstructed $m_{\tau\tau}$ mass of the events observed in various categories, while the ATLAS analysis fits the scores of Boosted Decision Trees (BDT's) trained to discriminate signal from background events.

The CMS event selection shows that most of the sensitivity comes from boosted events where the Higgs boson candidate, as reconstructed by the vectorial sum of the visible tau decay products and the missing transverse energy, has a transverse momentum greater than $p_{\mathrm T}^{\tau\tau}>100$ GeV. This selection defines the most sensitive event categories for both the VBF and the gluon fusion (ggF) productions. 
%The ATLAS analysis has no explicit selection on $p_{\mathrm T}^{\tau\tau}$ in the VBF event category, even though events with high score are likely to have high $p_{\mathrm T}^{\tau\tau}$ too. This is shown by a support analysis \cite{Aad:2015vsa}, similar to the CMS one.

Both ATLAS and CMS observed an excess of events over the background-only hypothesis with observed (expected) significance of 4.4 (3.3) and 3.4 (3.7) standard deviations, respectively. The measured Higgs boson production cross section times branching ratio in units of SM predictions (``signal strength'') is $\mu=1.41^{+0.40}_{-0.36}$ for ATLAS and $\mu=0.88^{+0.30}_{-0.28}$ for CMS. 

In combination with the ATLAS and CMS results in the other Higgs boson decays \cite{Khachatryan:2016vau}, these two analyses lead to the observation of the $H\to\tau\tau$ decays at $5.5\sigma$ ($5.0\sigma$) and $\mu=1.11^{+0.24}_{-0.22}$. From this combination, the signal strengths for the ggF and VBF productions measured in the $H\to\tau\tau$ decays are $1.0^{+0.6}_{-0.6}$ and $1.3^{+0.4}_{-0.4}$, respectively. The precisions on these results show that the observed sensitivity of the $H\to\tau\tau$ channel in Run-I is mostly from VBF events. In fact, in Run-I the $H\to\tau\tau$ channel has one of the best observed sensitivities to the VBF production, equal to the $H\to WW$ and better than the $H\to \gamma\gamma$ and $H\to ZZ$ channels.

\section{\cp Invariance in VBF Production with $H\to\tau\tau$ Events}
\label{sec:VBF}
The sensitivity of the VBF $H\to\tau\tau$ channel is exploited for a first direct test of the \cp invariance in the VBF Higgs boson production \cite{Aad:2016nal}. This is a test performed by ATLAS for \cp-violating contributions in the $HVV$ couplings, independently of the Higgs boson decay, in 20.3 fb$^{-1}$ of $pp$ collisions collected at $\sqrt{s}=8$ TeV. Investigations of \cp-violating Higgs boson couplings in the decays into pairs of massive gauge bosons show no deviation from SM \cite{Aad:2015mxa,Khachatryan:2014kca}.
The analysis uses the $H\to\tau\tau$ events selected in the VBF category as in Ref~\cite{Aad:2015vsa} in the final states where the tau leptons decay either both leptonically (\emph{ll}) in muons or electrons, or semi-leptonically (\emph{lh}) with one leptonically and one hadronically decaying tau lepton. 
%The event selection and the background estimation are as in Ref~\cite{Aad:2015vsa}. 
The events classified at high BDT-score are used in a fit to a \cp-sensitive observable, while the ones at low BDT-score are used to determine backgrounds.

\begin{figure}[tpb]
\begin{center}
\includegraphics[width=0.35\textwidth]{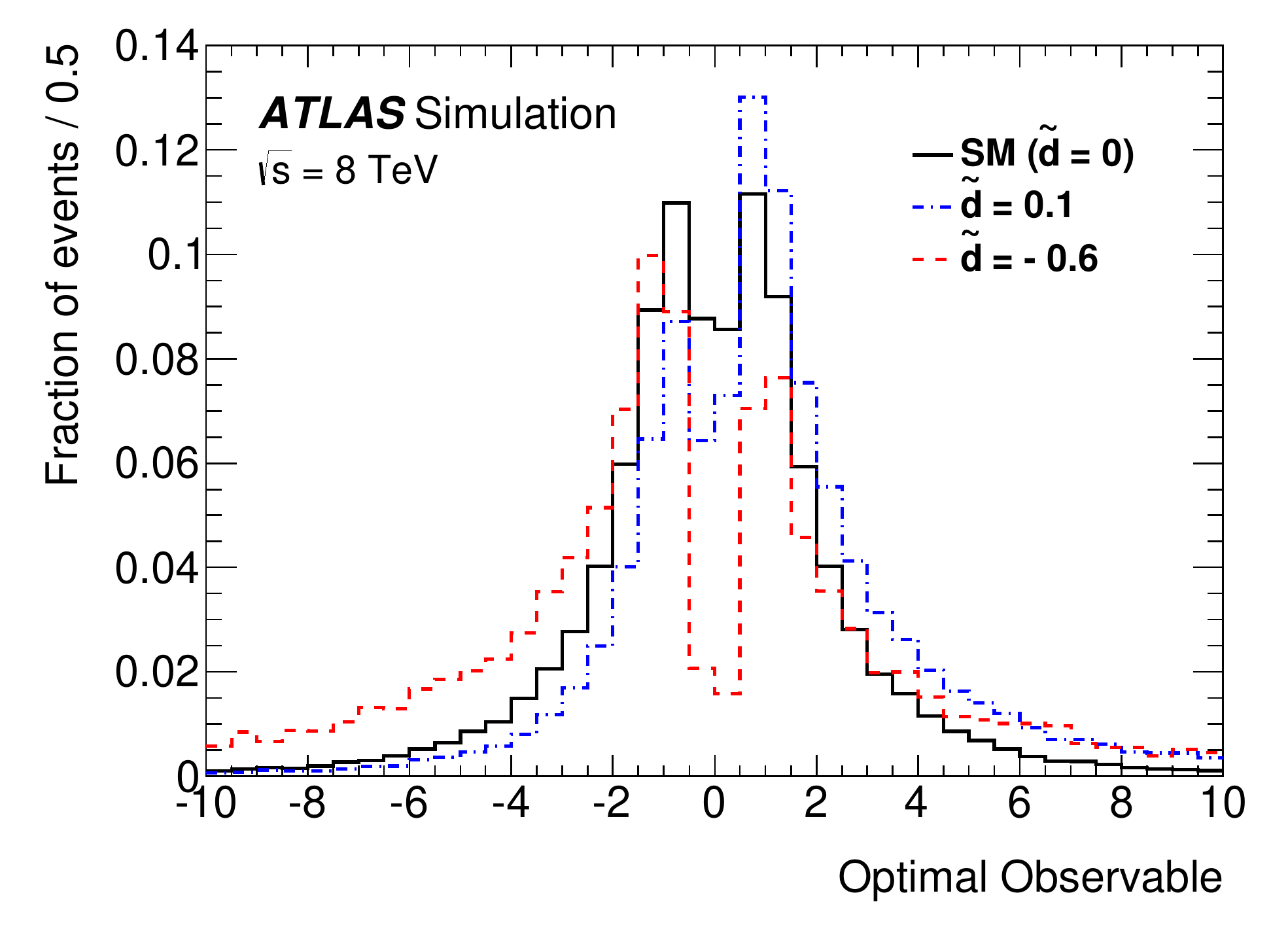}
\end{center}
\caption{Distributions of the \emph{Optimal Observable} for the SM hypothesis and two arbitrary values of $\tilde{d}$ in simulated signal events accepted in the VBF event category \cite{Aad:2016nal}.}
\label{fig:OO}
\end{figure}

The observable is a \cp-odd \emph{Optimal Observable} \cite{Atwood:1991ka,Davier:1992nw,Diehl:1993br} built from the leading-order matrix element for the VBF production. An effective Lagrangian is used to include \cp-violating effects from operators with mass dimension up to six in the $HVV$ couplings. The effective Lagrangian assumes that the same coefficient is multiplying the \cp-odd structures for the $HW^{+}W^{-}$, $HZZ$ and $H\gamma\gamma$ vertices and that all other couplings are as predicted in the SM. Under these assumptions, the matrix element $\mathcal{M}$ is the sum of the SM \cp-even contribution $\mathcal{M}_{\rm SM}$ and a \cp-odd contribution $\mathcal{M}_{{\rm cp-odd}}$ from the dimension-six operators parametrised by the parameter $\tilde{d}$:
\begin{equation}
\mathcal{M}=\mathcal{M}_{\rm SM}+\tilde{d}\cdot\mathcal{M}_{{\rm cp-odd}}.
\label{eq:M}
\end{equation}
The \emph{Optimal Observable} $\mathcal{OO}$ is defined as the ratio of the \cp-odd interference term between $\mathcal{M}_{\rm SM}$ and $\mathcal{M}_{{\rm cp-odd}}$ to the SM contribution:
\begin{equation}
\mathcal{OO}=\frac{2{\rm Re}(\mathcal{M}^{*}_{\rm SM}\mathcal{M}_{\rm cp-odd})}{|\mathcal{M}_{\rm SM}|^{2}}.
\label{eq:OO}
\end{equation}
Since the \emph{Optimal Observable} is \cp-odd, a non vanishing mean value $\langle\mathcal{OO}\rangle\ne0$ is an indication of \cp violation, as shown in Fig~\ref{fig:OO}.

\begin{figure}[tpb]
\begin{center}
\includegraphics[width=0.33\textwidth]{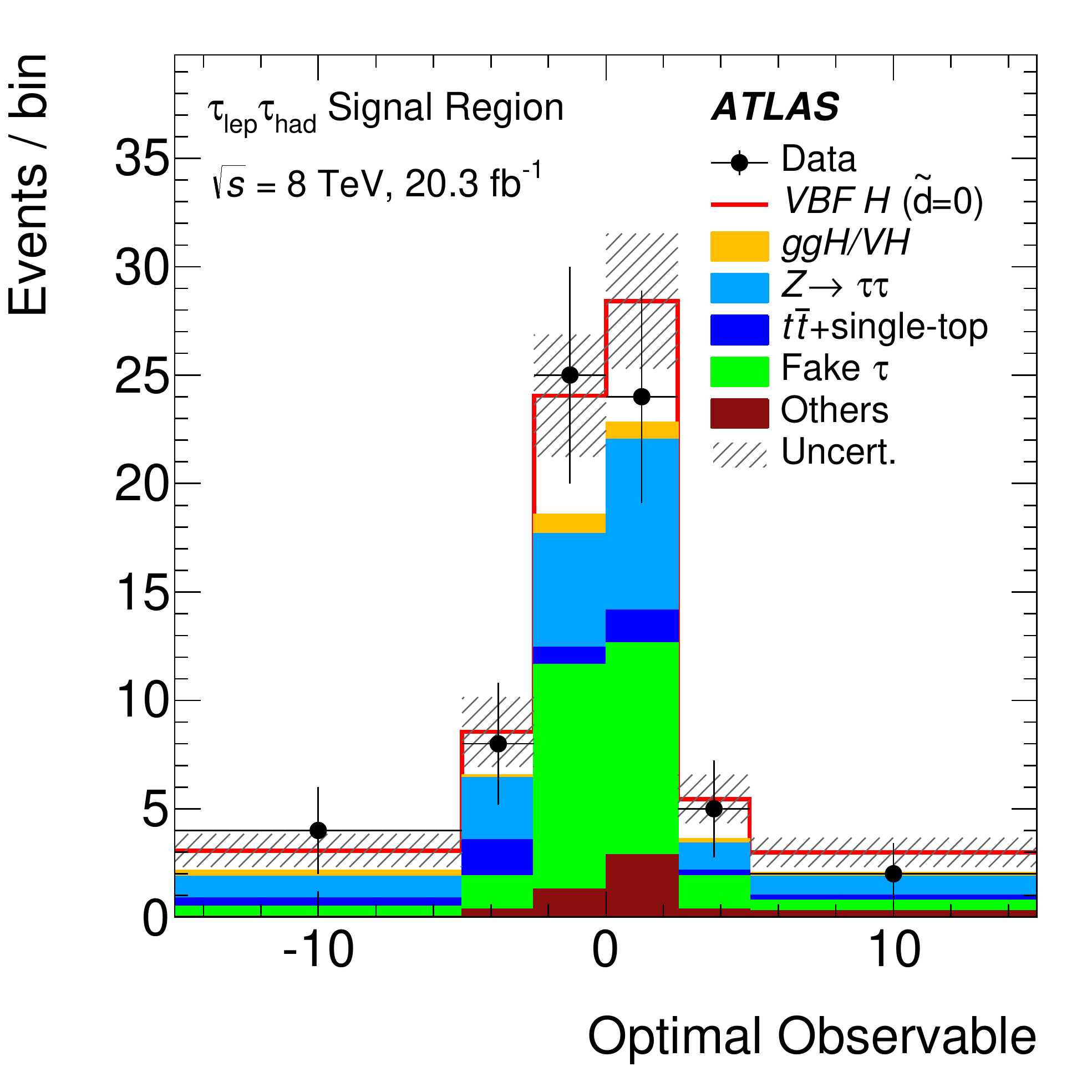}
\end{center}
\caption{Observed and expected distributions of the \emph{Optimal Observable} in the \emph{lh} signal region after the fit for the SM hypothesis. Signal is stacked on top of the background contributions with the best-fit $\mu=1.55^{+0.87}_{-0.76}$. Error bands include all uncertainties \cite{Aad:2016nal}.}
\label{fig:lh}
\end{figure}

The observed mean values $\langle\mathcal{OO}\rangle$ in the data selected in the signal regions are $0.3\pm0.5$ in the \emph{ll} channel and $-0.3\pm0.4$ in the \emph{lh} channel. Both results are consistent with zero within uncertainties and show no hints for \cp violation. Limits on the \cp-odd couplings are set based on a combined maximum-likelihood fit to the \emph{Optimal Observable} distributions in data both in the \emph{ll} and \emph{lh} channels. Fig~\ref{fig:lh} shows the result of this fit for the SM $\tilde{d}=0$ hypothesis and the best-fit $\mu=1.55^{+0.87}_{-0.76}$ in the \emph{lh} signal region. 
The regions $\tilde{d}<-0.11$ and $\tilde{d}>0.05$ are excluded at 68\% CL. These intervals are an order of magnitude better than those obtained by ATLAS using the Higgs boson decays into gauge bosons \cite{Aad:2015mxa}. The analysed data does not provide enough sensitivity to set 95\% CL intervals though.

\section{\cp Invariance in $H\to\tau\tau$ Decays}
\label{sec:decay}
The $H\to\tau\tau$ decays offer another unique chance to search for \cp violation in the Higgs sector and this is by means of probing the $H\tau\tau$ coupling. The $H\to\tau\tau$ is not only the most sensitive fermionic channel, but thanks to the maximally parity violating tau decays it also allows to access directly and in a model independent way the \cp structure of the Higgs boson Yukawa couplings. 

Unlike for the $HVV$ couplings, the SM Lagrangian can be extended at tree level with \cp-odd Yukawa couplings to allow for CP violation (see e.~g. Ref~\cite{Berge:2015nua}):
\begin{equation}
\mathcal{L}_{h\tau\tau}=-\frac{m_{\tau}}{v}\kappa_{\tau}(\cos\phi_{\tau}\bar{\tau}\tau + \sin\phi_{\tau}\bar{\tau}i\gamma_{5}\tau)h
\label{eq:Htt}
\end{equation}
where the relative contribution of the \cp-even and \cp-odd terms are parametrised by the mixing angle $\phi_{\tau}$. The SM coupling is realised by $\phi_{\tau}=0$, while the pure \cp-odd coupling corresponds to $\phi_{\tau}=\pi/2$. The mixing angle $\phi_{\tau}$ determines the transverse spin correlations between the two tau leptons and subsequently the angular distributions of the tau decay products. 

Because of the presence of at least two neutrinos in the final state, it is not possible to fully reconstruct the kinematics of the $H\to\tau\tau$ decays, that is the tau momenta and the Higgs rest frame where the sensitivity to $\phi_{\tau}$ would be maximal. However, among all channels the fully-hadronic (\emph{hh}), where both tau leptons decay hadronically, is expected to be the most sensitive because of the only two neutrinos produced.

Several observables based on angles between tau decay planes built from the visible parts of the tau decay have been proposed \cite{Berge:2015nua,Berge:2014sra,Berge:2013jra,Harnik:2013aja,Desch:2003rw,Bower:2002zx,Jozefowicz:2016kvz}. In case of direct decays $\tau^{\pm}\to\pi^{\pm}\nu$, the decay plane can be defined by the pion track and its impact parameter with respect to the point of the primary interaction \cite{Berge:2015nua,Berge:2014sra,Berge:2013jra}. This method can be extended to any other tau decay. Another method targets the $\tau^{\pm}\to\rho^{\pm}(770)\nu\to\pi^{\pm}\pi^{0}\nu$ decays and uses the tau decay plane spanned by the momenta of the two pions \cite{Desch:2003rw,Bower:2002zx}. In both methods, event classifications based on the pion energies are used to increase the sensitivity.

At present, there is no study on this measurement with full detector simulation, but an assessment of the impact of the tau decay reconstruction, especially for hadronic decays, is essential to get reliable sensitivity estimates.
 
\begin{figure}[tpb]
  \centering
\includegraphics[width=0.3\textwidth]{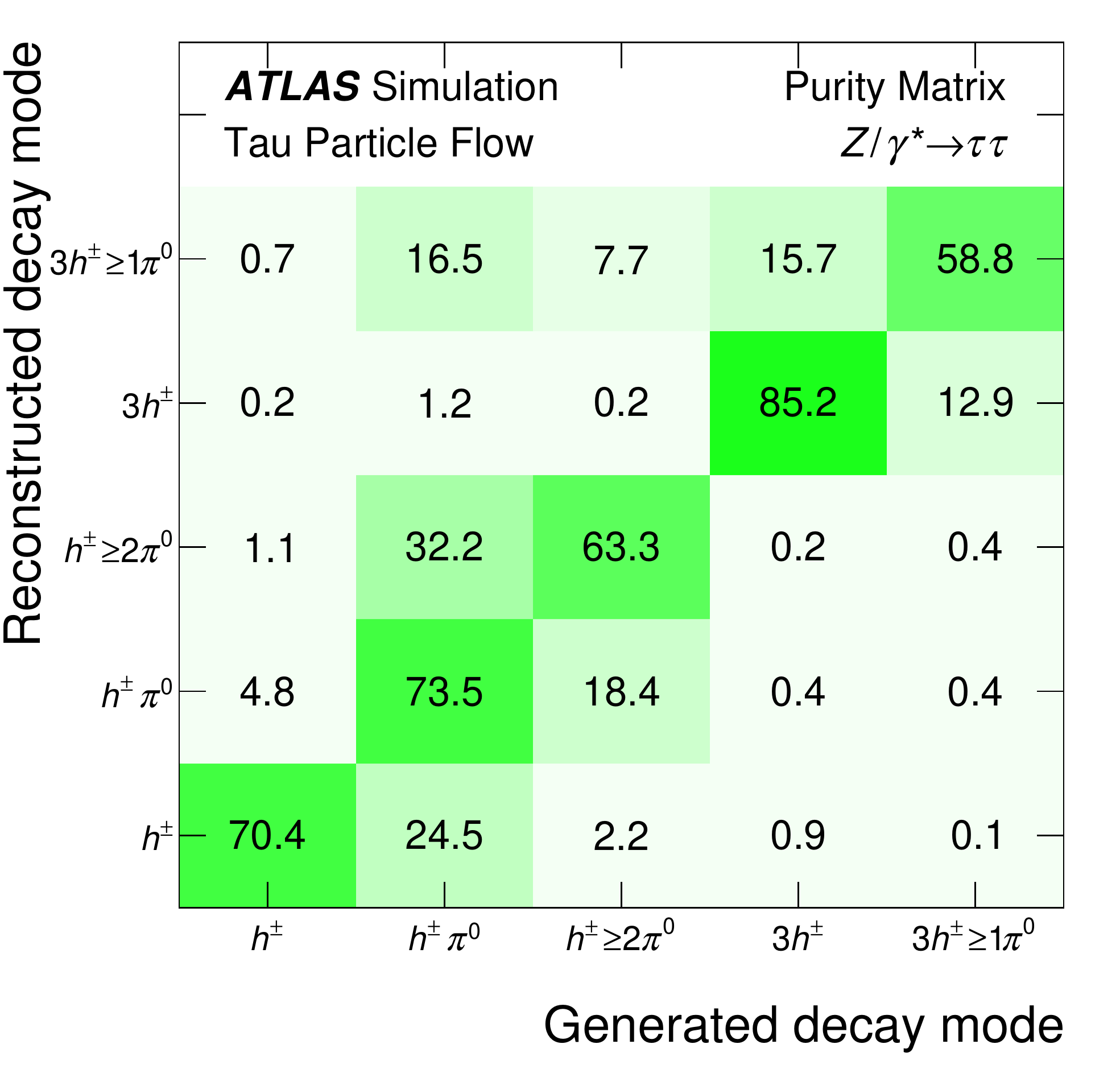}
\caption{Decay mode classification purity matrix showing the fraction of $\tau_{\rm had}$ candidates of a given reconstructed mode that originated from a generated $\tau_{\rm had}$, in $Z\to\tau\tau$ events simulated with LHC Run-I conditions \cite{Aad:2015unr}.}
\label{fig:purity}
\end{figure}

CMS and, since the LHC Run-II, also ATLAS use particle flow algorithms for the $\tau_{\rm had}$ reconstruction \cite{Aad:2015unr,Khachatryan:2015dfa} in which the charged components are reconstructed from tracks, while the neutral components are reconstructed from calorimeter clusters not matched to tracks. These algorithms are crucial for this measurement as they provide information on the tau decay products needed to classify the tau decay modes and to build the tau decay planes. The reconstruction of the neutral hadrons is the most challenging part of the $\tau_{\rm had}$ reconstruction and it is expected to have a large impact on the sensitivity of the measurement in terms of decay mode classification and decay plane resolution.
In ATLAS for example, the $\tau_{\rm had}$ reconstruction efficiency, including the detector acceptance, ranges between 32\% to 43\%, while the identification efficiency goes from 40\% to 75\%, with the tau decays with neutral components having the lowest identification 	efficiencies \cite{Aad:2015unr}. Among the reconstructed $\tau_{\rm had}$, 74.7\% of the decay modes are correctly classified. The modes with higher misclassification rates are those with neutral hadrons where one of the neutrals is not identified, as shown in Fig~\ref{fig:purity} \cite{Aad:2015unr}.

\begin{figure}[tpb]
  \centering
\includegraphics[width=0.35\textwidth]{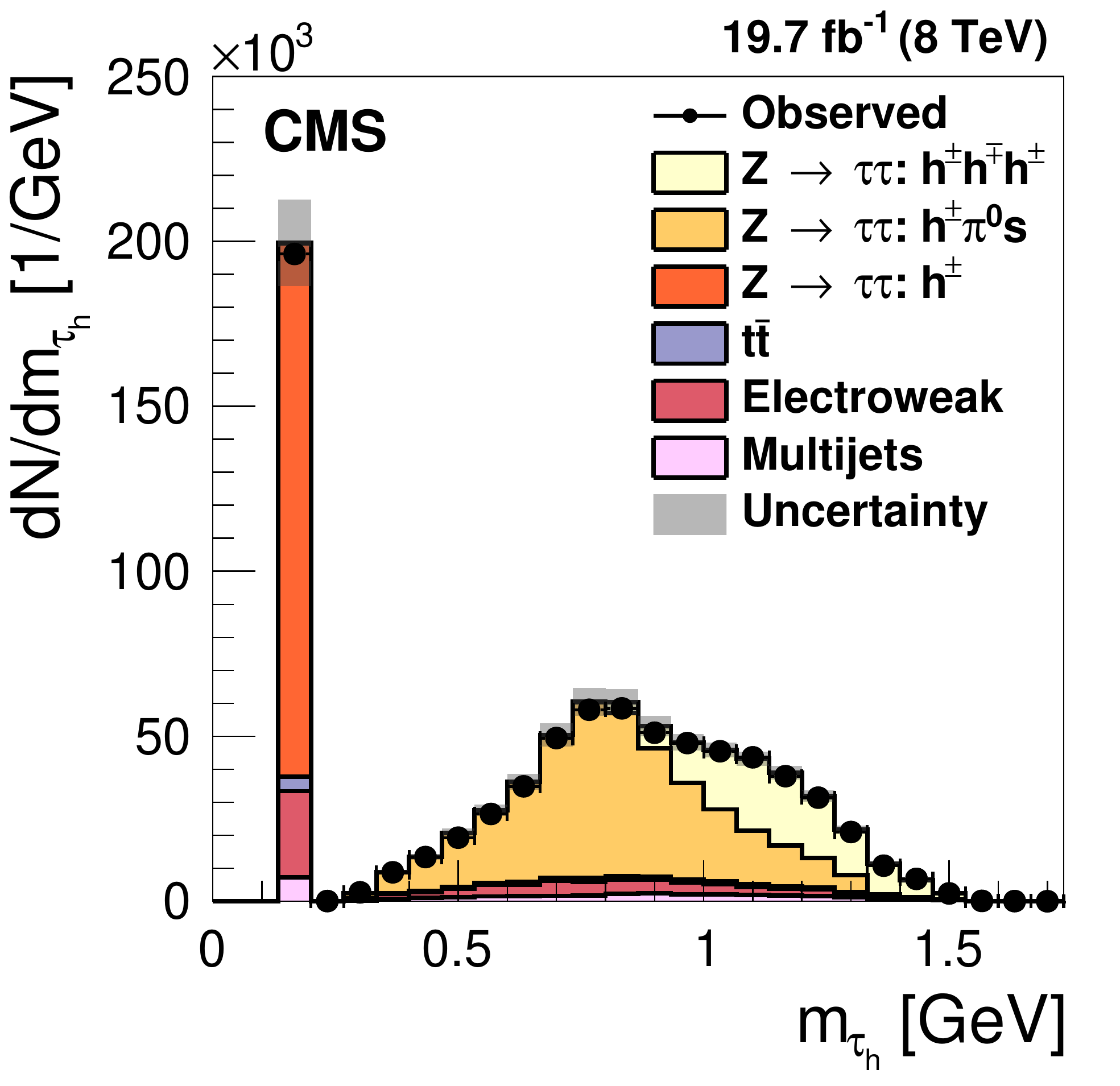}
\caption{Distribution of $\tau_{\rm had}$ candidate masses in $Z\to\tau\tau$ events selected in data, compared to MC expectations \cite{Khachatryan:2015dfa}.}
\label{fig:taumass}
\end{figure}

The angular resolution on the momenta of the neutral hadrons is at the per cent level, but the transverse energy is only at about 16\%.
Despite the challenging task, overall the ATLAS and CMS $\tau_{\rm had}$ reconstructions have good performances and impressive agreement between data and simulation, as shown for instance in Fig~\ref{fig:taumass} \cite{Khachatryan:2015dfa}.

Besides the $\tau_{\rm had}$ reconstruction, other challenges for the \cp violation measurement come from the trigger and the background estimation. The bottleneck for the signal acceptance in the \emph{hh} final state is the first trigger level, where only calorimeter information is available. Triggering on two low-energy calorimeter clusters is very expensive in terms of rate, so additional selections on an additional jet or on the $\Delta R(\tau_{\rm had},\tau_{\rm had})$ between the two tau candidates can be used to reduce rate while keeping high acceptance in the boosted phase space ($p_{\mathrm T}^{\tau\tau}\gtrsim100$ GeV) used for the analysis. 
%The more extensive use of tracking information in the triggers being developed by ATLAS and CMS for the High-Luminosity LHC will be very beneficial for this measurement.
The background estimation could also pose challenges. The dominant and irreducible $Z\to\tau\tau$ background is expected to be flat in the sensitive observables \cite{Berge:2014sra} since the tau leptons are mostly longitudinally polarised. However, it is important to validate the modelling of the tau spin correlations in the $Z$+jets events accepted in the analysis signal regions. If the spin correlations in these events are significantly different from those produced via Drell-Yan (DY) production, corrections or alternatives to the $Z\to\tau\tau$ modelling via embedded data $Z\to\mu\mu$ events \cite{Aad:2015vsa,Chatrchyan:2014nva, Aad:2015kxa} as used in the Run-I analyses may be needed. In fact, in the embedded events the simulated $Z\to\tau\tau$ decay always assumes the DY production. The modelling of the multi-jet background used in Run-I may also need to be revisited since it uses $\tau_{\rm had}$ reconstructed with multiplicities of charged and neutral hadrons different from the $\tau_{\rm had}$ candidates accepted in the signal regions.  This could lead to biases in the construction of the decay  planes.

\section{Conclusions}
\label{sec:conclusions}
The $H\to\tau\tau$ events have great potential for the direct measurement of \cp violation in the Higgs sector. Evidence for these events at 5.5$\sigma$ has been established combining the ATLAS and CMS Run-I analyses. This observation is driven by the detection of events at high $p^{\tau\tau}_{T}$ and in particular from the VBF production. The sensitivity of the VBF $H\to\tau\tau$ channel has been exploited in a first direct test of \cp invariance in the VBF production that leads to results ten times more stringent than previous ones from the diboson channels.
The direct and model independent measurement of \cp violation in the $H\to\tau\tau$ decays is also very promising. The outlined experimental challenges comprise the reconstruction from the neutral hadrons in the tau decay, the trigger strategy and the background estimates for $Z\to\tau\tau$ and multi-jet events.

%\begin{figure}[htbp]
%\begin{center}
%\includegraphics[width=0.35\textwidth]{HiggsDecay_phi-eps-converted-to.pdf}
%\end{center}
%\caption{}
%\label{fig:decayCP}
%\end{figure}
%
%\begin{table}[tbp]
%\centering
%\begin{tabular}{lccc}
%\hline
%Decay mode & $\mathcal{B}$ [\%] & $\mathcal{A}\cdot\varepsilon_{\rm reco}$ [\%] & $\varepsilon_{\rm ID}$ [\%] \\
%\hline
%$h^{\pm}$   &  11.5  & 32 & 75 \\
%$h^{\pm}\pi^{0}$    &  30.0  & 33 & 55 \\
%$h^{\pm}\ge2\pi^{0}$      &  10.6  & 43 & 40 \\
%$3h^{\pm}$ &   9.5  & 38 & 70 \\
%$3h^{\pm}\ge1\pi^{0}$    &   5.1  & 38 & 46 \\
%%
%\hline
%\end{tabular}
%\caption{
%%
%%
%}
%\label{tab:decaymodes}
%\end{table}

%\begin{figure}[htbp]
%\begin{center}
%\includegraphics[width=0.4\textwidth]{fig_03a.pdf}
%\end{center}
%\caption{}
%\label{fig:decayCP}
%\end{figure}

%\begin{figure}[htbp]
%\begin{center}
%\includegraphics[width=0.4\textwidth]{CMS-TAU-14-001_Figure_003-a.pdf}
%\includegraphics[width=0.4\textwidth]{fig_07a.pdf}
%\end{center}
%\caption{}
%\label{fig:decayCP}
%\end{figure}

%% The Appendices part is started with the command \appendix;
%% appendix sections are then done as normal sections
%% \appendix

%% \section{}
%% \label{}

%% References
%%
%% Following citation commands can be used in the body text:
%% Usage of \cite is as follows:
%%   \cite{key}         ==>>  [#]
%%   \cite[chap. 2]{key} ==>> [#, chap. 2]
%%

%% References with BibTeX database:
\nocite{*}
\bibliographystyle{elsarticle-num}
\bibliography{martin}

%% Authors are advised to use a BibTeX database file for their reference list.
%% The provided style file elsarticle-num.bst formats references in the required Procedia style

%% For references without a BibTeX database:

\end{document}